\documentclass[aps,prb,twocolumn,groupedaddress,showpacs,showkeys,floatfix]{revtex4}

\usepackage{amsmath}
\usepackage{amsfonts}
\usepackage{amssymb}
\usepackage{graphics}
\usepackage{graphicx}

\begin{document}

%\preprint{}

\title{Localization and hybridization of
electronic states in thin films of Ag on V(100)
}

\author{P.~Lazi\' c}
\author{\v Z.~Crljen}
\author{R.~Brako}
\email[]{radovan@thphys.irb.hr}
%\thanks{}
\affiliation{R. Bo\v skovi\' c Institute, P.O. Box 180, 10002 Zagreb,
     Croatia}

\date{\today}

\begin{abstract}
We have studied the electronic states in 1--5 layers thick
Ag films on V(100), by means of \textit{ab initio} density functional
calculations. Due to the mismatch of the electronic structure of Ag and V,
quantum well states of both sp and d character localized on Ag films are formed.
We find that the hybridization of the Ag quantum well states with the V orbitals is 
nevertheless important, and must be taken into account in order to fully 
understand the observed properties, in particular the energies and
the dispersion of the photoemission peaks in ARPES experiments.
\end{abstract}

%\keywords{ }

\maketitle

\section{\label{introduction}Introduction}

Ultrathin metallic films grown on a different metal substrate show 
interesting structural and electronic properties. Some grow in 
registry with the substrate, while others form a variety of commensurate
or incommensurate structures, develop grain boundaries, etc. Of particular
interest are the systems which exhibit quantum size effects (QSE), where
the electronic states in the films, with thickness typically from one
up to several tens of layers, show quantized motion in the
direction perpendicular to the surface plane. Quantum well (QW) states 
or quantum well resonances  can be formed, depending on the degree
of mismatch between the electronic structure of the two metals, and the
properties can vary a lot over the extent of the two-dimensional
surface Brillouin zone (BZ).

Several experimental techniques can be used to probe the electronic properties 
of thin metallic films, and in particular the quantum well states. 
For an overview see, e.g., a recent review paper.\cite{Mil-02} 
Most commonly used are angle-resolved photoemission spectroscopy (ARPES)
and the scanning tunneling microscopy (STM). ARPES has been used mostly
in the normal emission mode, i.e.\ in the direction perpendicular to the
surface, but angularly resolved spectra in off-normal directions make it
possible to study the band structure
along any direction of the surface Brillouin zone.\cite{Joh-94, Mat-02} 
The photoemission spectroscopy probes a macroscopic
surface region, determined by the size of the incoming photon beam, which
means it is  best suited for systems of uniform thickness. A well known
example of this kind are silver films on a Fe(100) whisker, which can be
grown up to 120 macroscopically uniform atomic layers.\cite{Pag-99}
STM is  a nanoscopically local method, and offers the possibility to 
study local structures, such as islands, chains, dots, etc. With the
scanning tunneling spectroscopy (STS) technique, in which the
differential conductance ($dI/dV$) is measured, one can reliably determine the 
energy of quantized electronic states in the range of approximately 1 eV below and
above the Fermi level. The constant current STM technique can also give
important information about the surface electronic structure, in particular
by direct imaging of standing waves formed around surface defects in the
low bias voltage (and usually low temperature) regime, and taking the
Fourier transform. This method is known as the Fourier transform scanning 
tunneling microscopy (FT-STM).\cite{Spr-97}

In this paper we study the properties of thin films of silver on
a V(100) surface. The films can be grown pseudomorphically, i.e.\ in registry with the
substrate, up to a thickness of at least ten layers. It has been recently shown 
that, by using an appropriate deposition and annealing procedure, films of 
very high quality can be obtained even on V(100) surface with oxygen-induced 
($5\times1$) reconstruction, because silver atoms displace oxygen and 
lift the reconstruction.\cite{Kra-03a} The silver films grow as (111) layers
of fcc structure, slightly tetragonally distorted due to a small 
mismatch of interatomic distances of the two lattices. The electronic properties of 
this system are very interesting, due to the profound difference of the
band structure of the two metals. In bulk vanadium, the s bands extend
approximately from 2.5 to 7 eV below the Fermi level, while the d bands
exist between the Fermi level down to 3 eV. The order is reverse in
bulk silver which has d bands between 3.5 and 6 eV below the Fermi level, 
and the broad sp bands extend from the Fermi level down to 7 eV. 
This symmetry mismatch is particularly important in the center of the
two-dimensional Brillouin zone of the V--Ag~(100) interface, where it amounts
to a {\it symmetry gap} which prevents hybridization of certain 
electron bands across the interface. In ultrathin Ag films on V(100),
it has a consequence that both the sp orbitals of Ag close to the
Fermi level and d orbitals at larger binding energies form QW states
which are mostly localized within the Ag layers. 
In normal ARPES, which selects electronic states at
the center of the two-dimensional Brillouin zone, 
the QW states appear as strong narrow peaks.\cite{Val-96, Kra-03b} 

A number of approaches have been used in the past in order to
describe the electronic properties, in particular the QW states,
in ultrathin metallic films. Quasi-one-dimensional models, such as 
a square well potential \cite{Kra-01} or the phase accumulation model \cite{Lin-87}
have been successfully  used to to interpret the energies of QW states.
More sophisticated methods have also been used, such as the tight-binding
approach \cite{Smi-94} and layer-Korringa-Kohn-Rostoker approach.\cite{Ern-02} 
In a few systems the QW states have been investigated
by self consistent density functional calculations.\cite{Car-00,Wei-03} 
There are strong reasons to use the \textit{ab initio} methods. First, there are no
adjustable parameters, and a wide range of calculated structural and 
electronic properties offer the possibility of a detailed 
comparison with experiments. Also, quantities such as the expected
STM profiles and the amplitudes of the wavefunctions of the QW states, which 
cannot be obtained in simple approaches, can be calculated.

In a recent paper we have reported our \textit{ab initio} calculations of the 
structure of the Ag/V(100) system.\cite{Kra-03a} We used the density functional 
theory (DFT) approach and calculated the structure and the total energy of 
silver films on the V(100) surface from one to five layers thick.
The calculated interlayer distances, step heights, work functions and 
other quantities were in excellent agreement with the experimental 
findings. In this work we use the same method
to calculate the {\it electronic} properties of the Ag/V(100) system and 
compare the results with photoemission experiments.

\section{\label{studies}
Experimental studies of A\lowercase{g}/V(100) films}

STM experiments give important information on 
structural properties of Ag/V(100) films, 
e.g. the disappearance
of the oxygen-assisted ($1\times5$) reconstruction of the bare V surface
upon silver adsorption and annealing, 
the step heights between Ag terraces of different number of layers, etc. 
In STM images the Ag terraces appear smooth, with small corrugation of
square symmetry corresponding to the atomic structure. 
In a recent experiment standing waves  
around impurities \cite{Kra-03c,Kra-05,Kra-04} have been observed, which 
indicates that the motion of electrons with energies around
the Fermi level is largely two-dimensional, confined to the 
Ag films. This is one of the rare examples where the 
oscillatory wave patterns at metal surfaces 
have been observed at room temperature.

In photoemission spectra along the surface normal
the quantum well states of predominantly sp character appear
as strong well defined peaks in the energy region between the Fermi
level and 2.5 eV below it. They are narrow and the position of the
maximum does not depend on the energy of the incident light,
which means that they are localized on the silver films. 
The peak position is a characteristic of the number of
Ag layers, and the appearance of one or more peaks in the 
photoemission spectra can be used to determine the thickness 
and the degree of perfection of the films. 
Off-normal photoemission measurements make it possible to 
determine the dispersion relation of the QW states.\cite{Huf-95}
Away from the center of the surface Brillouin zone, i.e.\ at 
finite $k_{||}$, the photoemission peaks of the QW states
broaden and the dependence upon $k_{||}$ 
often deviates from the parabolic free-electron behavior. 
Several of the theoretical approaches mentioned in the Introduction
have been used to treat the sp QW states of Ag/V(100) films,
including a simple square potential,\cite{Mil-99}
the phase accumulation model,\cite{Mil-99} the layer-Korringa-Kohn-Rostoker 
approach,\cite{Ern-02} and the tight-binding calculations.\cite{Kra-03b}
These calculations have made it possible to assign the 
experimentally observed
QW states to different families according to the
number of nodes of the wavefunction, for up to eight layers or so.\cite{Ern-02} 

The energy region from 2.5 to 7 eV below the Fermi level in
photoemission is dominated by features derived from d bands of Ag.
A number of sharp peaks are observed in normal photoemission, 
implying that some d orbitals form QW states localized
within silver films.

\begin{figure*}
\rotatebox{0}{
\resizebox{1.4\columnwidth}{!}{
\includegraphics[clip=true]{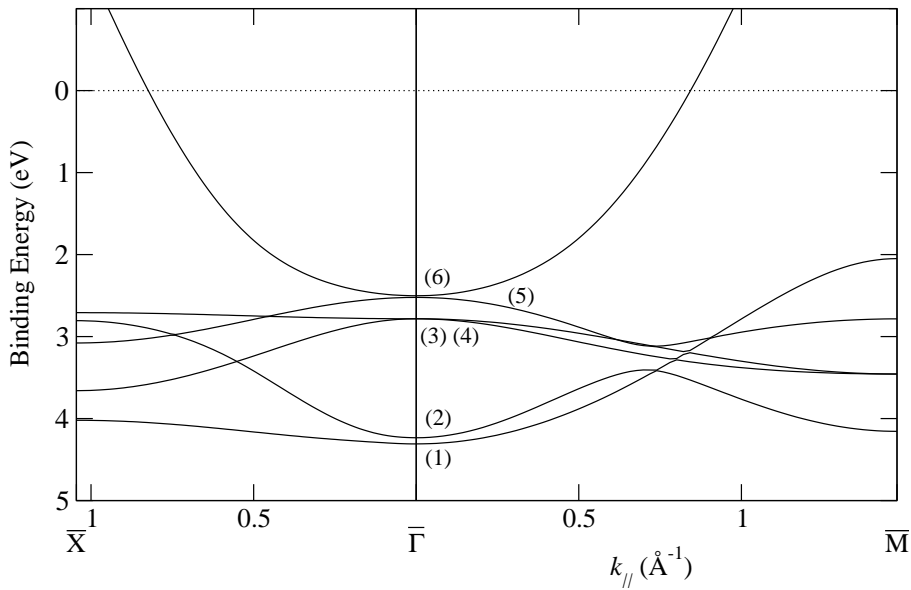}
}}
\caption{\label{monolayer}
The energies of electronic states of an unsupported Ag monolayer 
of square symmetry, with lattice constant corresponding to 
interatomic distance on a V(100) surface. The energies are calculated 
using DFT, in the high symmetry directions of the surface Brillouin
zone. In the center of the Brillouin
zone the character of the states is, starting from the lowest energy,
(1) $\textrm{d}_{x^2-y^2}$, (2) $\textrm{d}_{zz}$ with some 
s admixture, (3), (4) degenerate $\textrm{d}_{xz}$ and $\textrm{d}_{yz}$, 
(5) $\textrm{d}_{xy}$, and (6) predominantly s.
}
\end{figure*}

\begin{figure*}
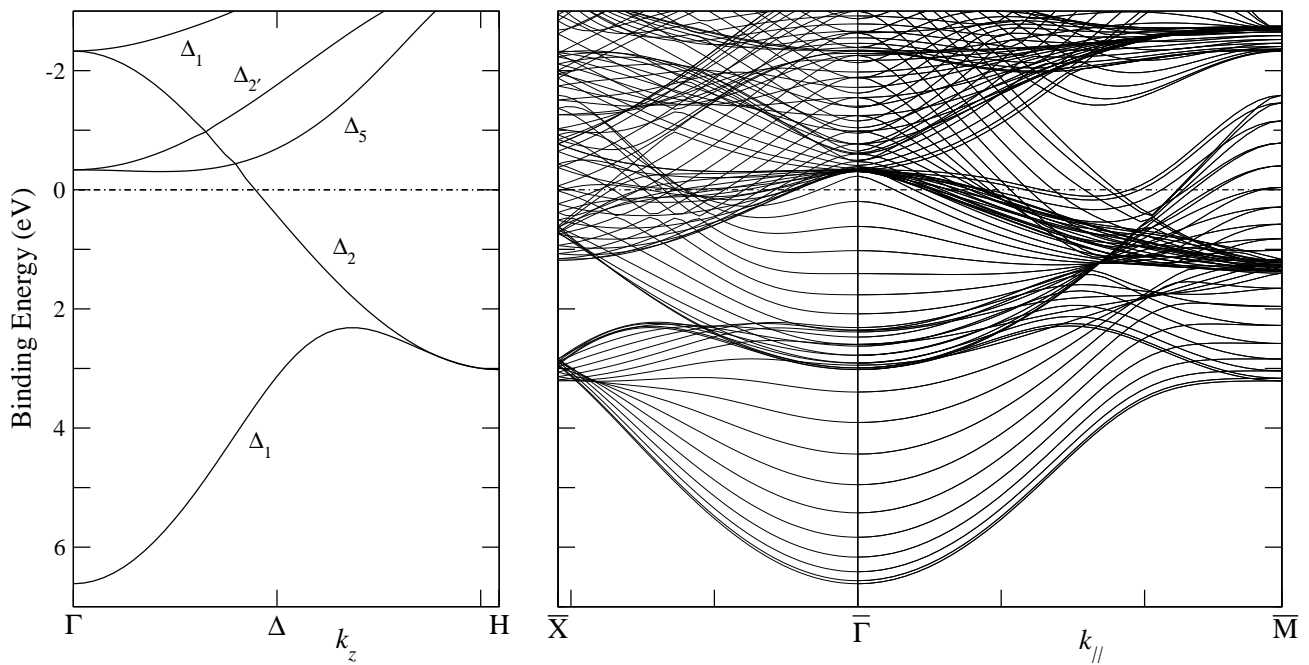

\rotatebox{0}{
\resizebox{2.0\columnwidth}{!}{
\includegraphics[clip=true]{Fig2a-V_bands-kz.eps}
\hspace{2em}
\includegraphics[clip=true]{Fig2b-V_bands.eps}
}}
\caption{\label{Vbulk}
The electronic structure of bulk vanadium. (a) In the direction
normal to the V(100) surface. The bands are labeled by their group 
representation. (b) Projection along high-symmetry directions of 
the (100) surface Brillouin zone, for 20 values of $k_z$.
}
\end{figure*}

\section{\label{calculation}Calculations}

The density functional calculations were performed using the DACAPO
program with ultrasoft pseudopotentials for the Perdew-Wang 
exchange-correlation functional, in the generalized gradient 
approximation (GGA).  The computational procedure has been described 
in considerable details in Ref.~\onlinecite{Kra-03a}, and here we only give a 
brief overview. Seven layers of vanadium were used to describe the
bcc V(100) substrate, on which one to five silver layers were added.
Since periodic boundary conditions were used in all three spatial
directions, a layer of vacuum of around 15~{\AA} was added on top of silver.
The structure was allowed to fully relax in the direction perpendicular
to the surface, except for the three bottom V layers, which were kept
fixed at the bulk interatomic distance. We used a mesh of $12\times12$ $k$ points
in the directions along the surface plane. Once the full selfconsistency 
of the relaxed structure was achieved, we made additional calculations 
with many more $k$ points (typically 36 from the center to the edge of the BZ) 
along the two high-symmetry
directions of the surface Brillouin zone, in order to obtain the
dispersion relations and the wavefunctions of the Kohn-Sham eigenstates
in more detail. Since in these calculations the
$k$ points do not cover uniformly the whole two-dimensional Brillouin zone,
the potential determined beforehand in selfconsistent calculations is used.

We use the calculated electronic eigenstate energies and wavefunctions 
for comparison with experiments, in particular photoemission. 
However, the Kohn-Sham density functional approach is designed primarily for
an accurate calculation of the total energy of the system. 
By varying the positions of the atoms one can determine the equilibrium
atomic configuration, the interatomic potential around the equilibrium, etc.
The electronic eigenstates are obtained as an intermediate result and
their quality and suitability for the description of various electronic
processes is not guaranteed.
The method is particularly unreliable for empty
electronic states above the Fermi level, which do not contribute towards
the total energy of the system and are therefore not included in the energy 
functional.
In fact, it is possible to perform the DFT calculations using just enough 
orbitals to accommodate all the electrons, at the expense of the rate of 
convergence, in which case the empty states will be obviously inadequate.
Even if sufficiently many orbitals are included, the vacuum layer separating 
the films in the direction perpendicular to the 
surface may not be large enough to prevent the 
coupling of states above the Fermi level due to a finite probability 
of tunneling through the vacuum.
Care should be exerted if empty states are used e.g. in the analysis of STM images.
Nevertheless, it has repeatedly been found that, 
if adequate precaution is used, the calculated electronic eigenstates yield
a good description of the electronic structure of the solid. 
From comparison with other calculations we have concluded 
that our calculation indeed gives good electronic structure, apart 
from some d bands of silver being too close to the Fermi level, which will be 
discussed in detail later on. Apparently, the d bands have been a problem in a 
number of DFT calculations of the electronic structure of noble metals.

\begin{figure*}
\rotatebox{0}{
\resizebox{1.4\columnwidth}{!}{
\includegraphics[clip=true]{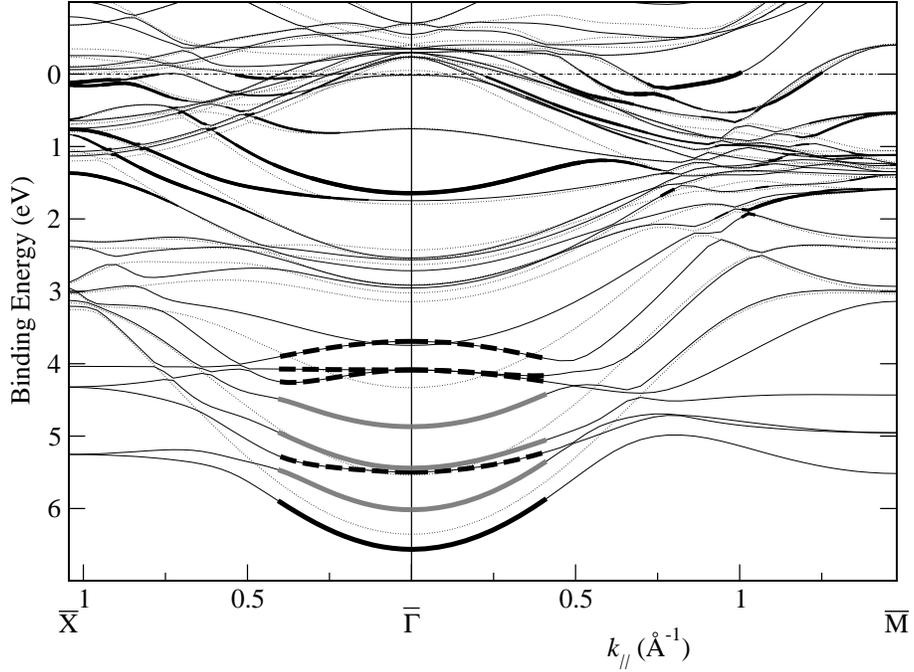}
}}
\caption{\label{1layer}
Similar as Fig.~\ref{monolayer}, but for a monolayer of Ag on seven layers
of vanadium (100) (full lines), and seven layers of V only (dotted lines). 
Around the center of the Brillouin zone we have denoted by thicker lines
the states of the Ag/V(100) system which have a large 
amplitude on the silver layer. Black lines:
states with s character, grey lines: Ag states of $\textrm{d}_{zz}$
symmetry, which hybridize strongly with the s band of vanadium,
dashed lines: Ag states of other d symmetries. In the region between
1.65 eV binding energy and the Fermi level we have calculated the
magnitude of various states on the silver layer, and represented it
by lines of different thickness. For a detailed discussion see text.
}
\end{figure*}

\begin{figure*}
\rotatebox{0}{
\resizebox{1.4\columnwidth}{!}{
\includegraphics[clip=true]{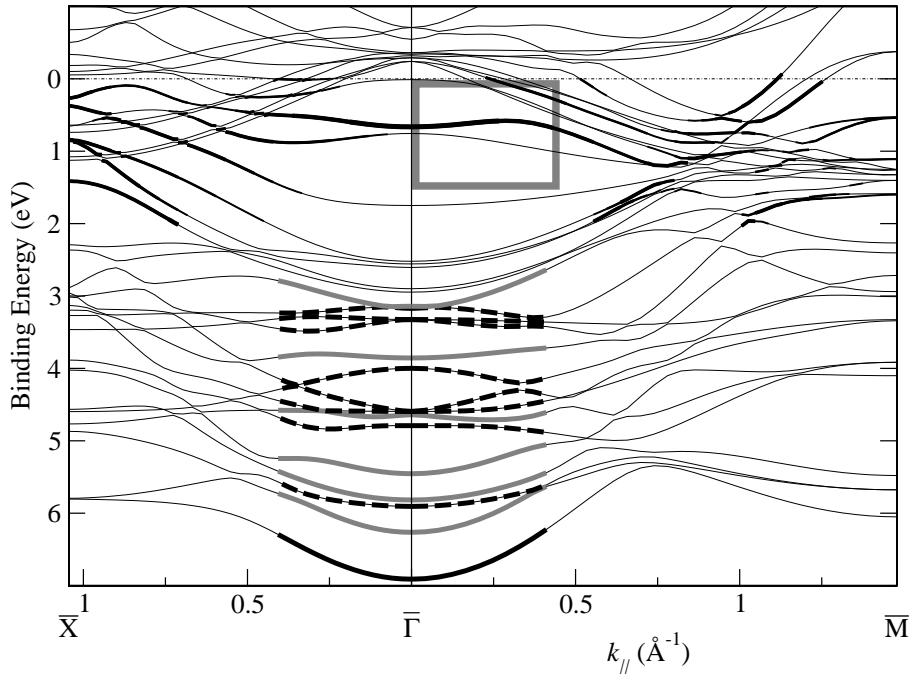}
}}
\caption{\label{2layers}
Similar as Fig.~\ref{1layer}, but for two layers of Ag on seven layers
of vanadium. The grey box denotes the $E-k_{\\}$ region which
has been investigated experimentally by off-normal ARPES in
Refs.~\onlinecite{Kra-03c,Kra-05}.
}
\end{figure*}

\section{\label{results}Results and comparison with experiment}

\subsection{The Kohn-Sham eigenstates}

We first made density functional calculations of an unsupported
Ag monolayer (i.e.\ a layer of silver atoms surrounded on both
sides by vacuum) of the same geometry as Ag atoms in a 1ML film
adsorbed on V(100). The energies of the Kohn-Sham eigenstates
are shown in Fig.~\ref{monolayer}, along the high symmetry directions 
of the surface Brillouin zone $\bar{\Gamma}$--$\bar{X}$ and 
$\bar{\Gamma}$--$\bar{M}$ 
The states between 2.5 and 4.5~eV binding energy with relatively
flat dispersion curves are d bands, while the s band shows 
a free-electron like dispersion and crosses the Fermi level at around
$k_{||}=0.8$~\AA$^{-1}$ in both calculated directions. Around the
center of the Brillouin zone the s orbitals also contribute some weight to
the $\textrm{d}_{zz}$ band, second from below, which is possible 
as both belong to the same completely symmetric representation of the
symmetry group at $\bar{\Gamma}$ point.

Next we calculated the Kohn-Sham energy bands of bulk vanadium and
projected them onto the directions relevant for the surface
calculations. In Fig.\ref{Vbulk}~(a) are shown the bands along the
$\Gamma$--$\Delta$--H direction, i.e. the $k_z$ direction in our surface
calculations, which projects onto the $\bar{\Gamma}$
point of the (100) vanadium surface Brillouin zone. In Fig.\ref{Vbulk}~(b)
we show the bands in the high symmetry directions of the surface Brillouin
zone, for 20 values of $k_z$. There are no true energy gaps at or near the 
$\bar{\Gamma}$ point. However, according to Fig.\ref{Vbulk}~(a) in the 
energy region between 2.3~eV and the Fermi level only states of
$\Delta_2$ symmetry exist. Consequently, the electronic states of 
silver adlayers at $\bar{\Gamma}$ which are totally rotationally 
symmetric with respect to the $z$ axis (i.e. s and $\textrm{d}_{zz}$)
which fall in this energy range 
cannot hybridize with the substrate bands. This is the so called
{\it symmetry gap}
which allows the existence of quantum well
states in adsorbate layers even when the substrate has no true
energy gaps. Of course, the symmetry incompatibility ceases to
be exactly true as soon as we move away from the $\bar{\Gamma}$ point.
Thus we expect that the photoemission peaks from such adsorbate QW
become broader when measured away from the normal direction, i.e.
for a finite $k_{||}$. This is unlike the QW states on substrates
with a true energy gap, where narrow photoemission peaks can be observed
for large $k_{||}$, until they hit either the Fermi level or 
the edge of a projected band.

At binding energies larger than approximately 3~eV the converse is
true. According to Fig.\ref{Vbulk}, only electronic states of the 
$\Delta_1$ symmetry exist at $\bar{\Gamma}$ point, which implies that 
there is a symmetry gap for d states other 
than $\textrm{d}_{zz}$ and the respective QW states may occur.

Finally, we come to the results for the combined vanadium--silver systems.
The dispersion relations of the eigenstates for 1 ML and 2 ML Ag films on
seven layers of V(100), calculated as explained in section 
\ref{calculation}, are shown in Figs. \ref{1layer} and \ref{2layers}. 
The thin dotted lines in Fig.~\ref{1layer} are the
results for bare vanadium substrate consisting of seven layers,
which are similar to the bulk dispersion relations projected onto
the surface, Fig.\ref{Vbulk}~(b), but more coarsely spaced because of 
the small number of layers.
The full lines are the results of the calculation
with the silver adlayer. It is easy to recognize
the electronic states which are localized mostly in vanadium, as
they run very close to the dotted lines of the clean V, while the
states associated with silver do not have a corresponding dotted
line. The states which at $k_{||}=0$ have a large weight on silver atoms
are shown by thick lines around the center of the Brillouin zone.
The different line styles denote states of different symmetry,
which has been deduced from the density of states projected onto 
atomic orbitals. In the energy range between the Fermi level 
and 2~eV we have gone a step further, and calculated the 
integral over the silver atoms of the density of various eigenstates, 
for the full range of $k_{||}$ vectors. The results are shown as
lines of various thickness, the thick segments corresponding to
states localized at silver atoms by more than around 35\%. The 
reasoning behind this approach
is that the calculated integrals are roughly proportional
to the intensity of the photoemission peaks in ARPES experiments,
since photoelectrons originating from deeper layers have a 
small probability of leaving the solid without being scattered.
The most prominent among the thick lines are the quantum well
states with energy 
of 1.65~eV for 1 ML and 0.66~eV
for 2 ML films at $\bar{\Gamma}$ point. 
They appear as strong narrow peaks in ARPES experiments, 
as discussed in the following subsection.

Comparing the states of 1~ML Ag films on vanadium, Fig.~\ref{1layer}, 
with the states of the unsupported Ag monolayer, 
Fig.~\ref{monolayer}, it is evident that the Ag d bands (except $\textrm{d}_{zz}$), drawn by
dashed lines, are pulled down to lower energies, while the shape of
the dispersion curves remains quite similar. This can be
attributed to the fact
that the effective potential well confining the states within the
Ag film is wider in the Ag/V(100) case than in the (rather
unrealistic) self-standing Ag film, but otherwise
the d states do not hybridize significantly with the vanadium
substrate. The situation is very different in the case of 
sp and $\textrm{d}_{zz}$
derived bands, which fall largely within the energy range of the 
vanadium s band and hybridize strongly with it. 
Around the center of the surface Brillouin zone the weight of the
Ag sp orbitals is transferred mostly to the quantum well state at
binding energy 1.65~eV mentioned in the preceding paragraph, 
and to the state at 6.56~eV, just below the
lower edge of the vanadium s band. The $\textrm{d}_{zz}$ state
(with some contribution of Ag sp orbitals) 
hybridizes strongly with the vanadium s band and contributes to several 
states, shown as thick grey lines. Their number would increase if we 
used more than 7 layers of V, while their weight on Ag atoms would 
decrease. Finally, in the limit of a thick vanadium substrate a 
broad Ag~$\textrm{d}_{zz}$--V~s resonance would appear.

A qualitative interpretation of the band structure of 2~ML Ag films, 
Fig.~\ref{2layers}, compared to the 1~ML results is that each d state 
localized on silver splits into two.
The sp-derived quantum well state with highest energy, 
well localized on Ag, is now at 
0.66~eV below the Fermi level. 

\begin{figure*}
\rotatebox{0}{
\resizebox{1.6\columnwidth}{!}{
\includegraphics[clip=true]{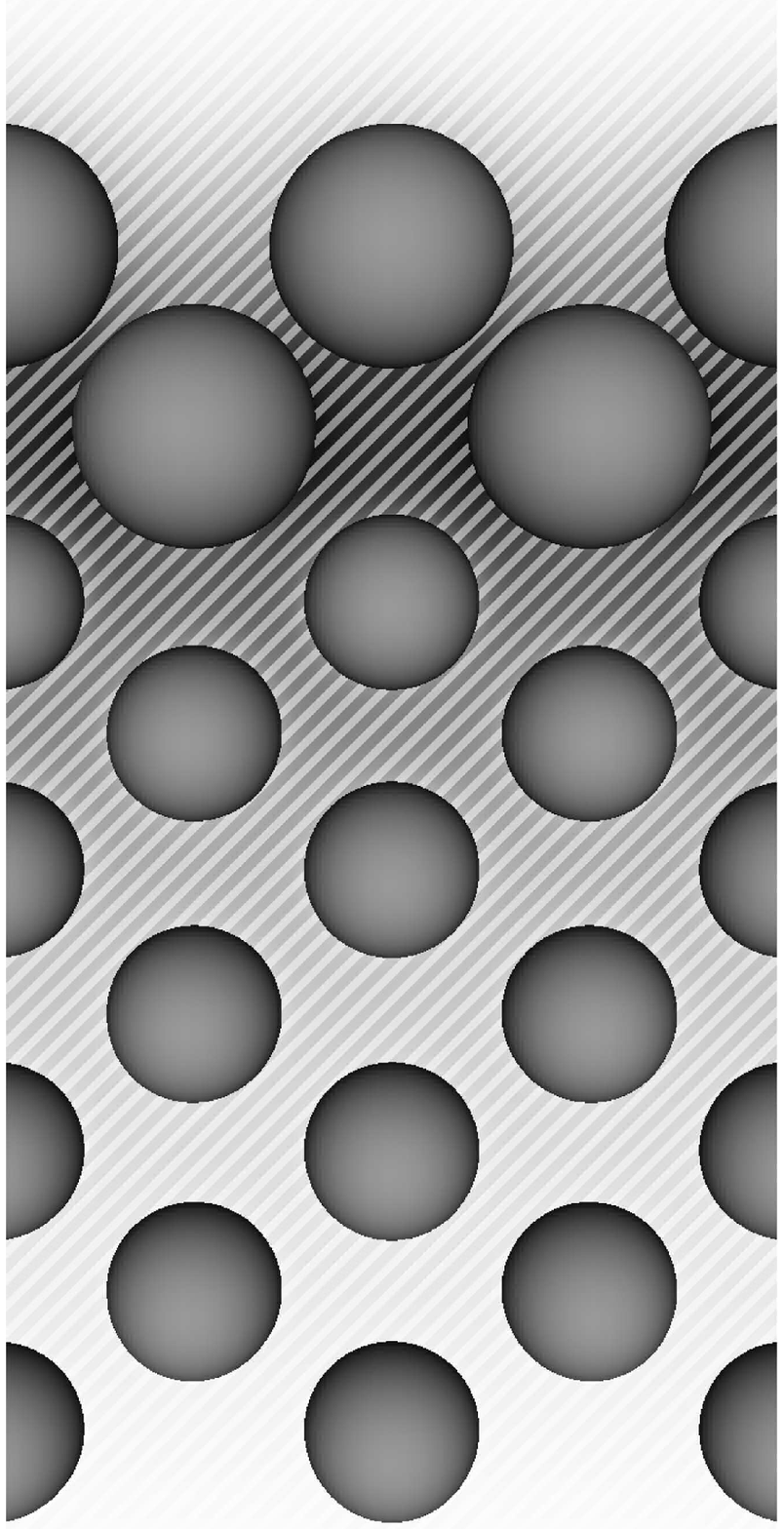}
\hspace{2em}
\includegraphics[clip=true]{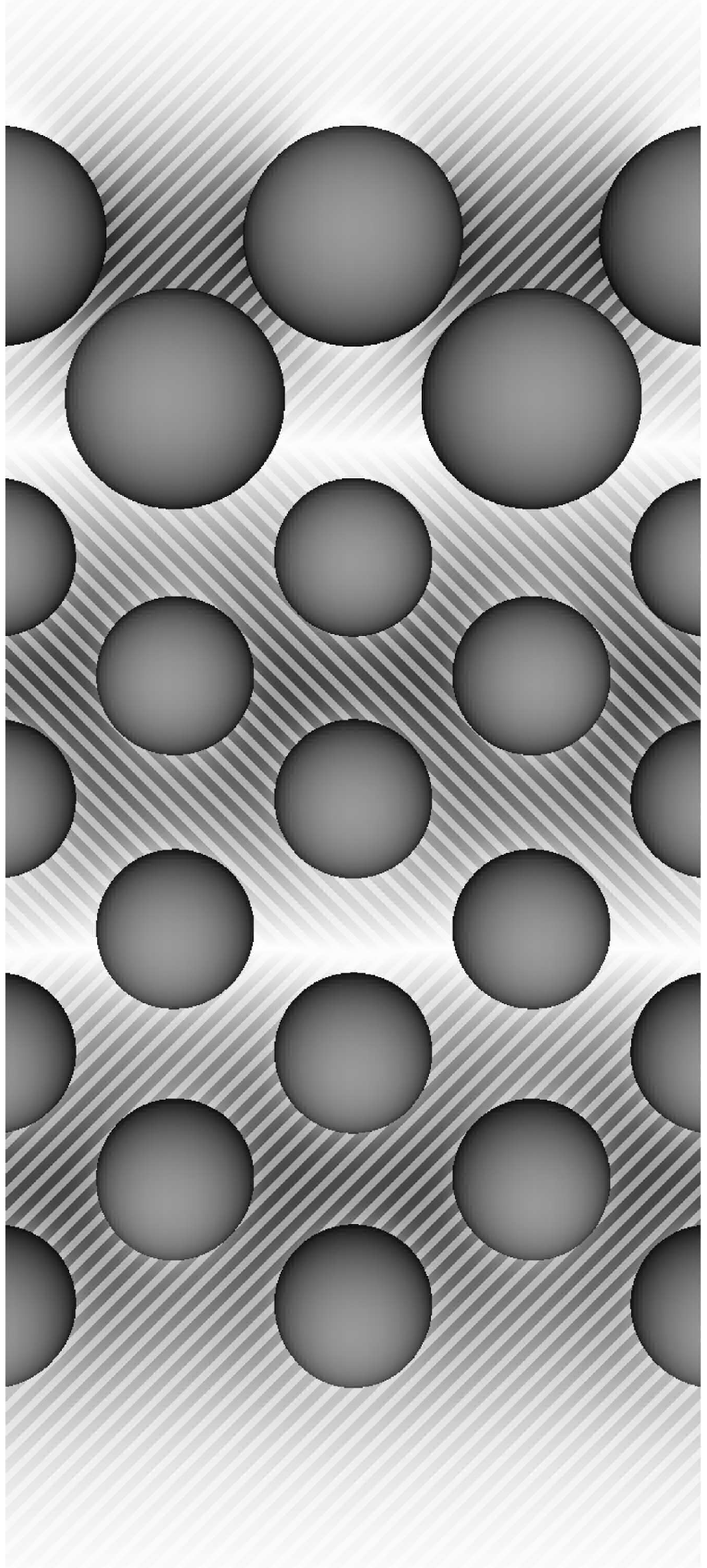}
\hspace{2em}
\includegraphics[clip=true]{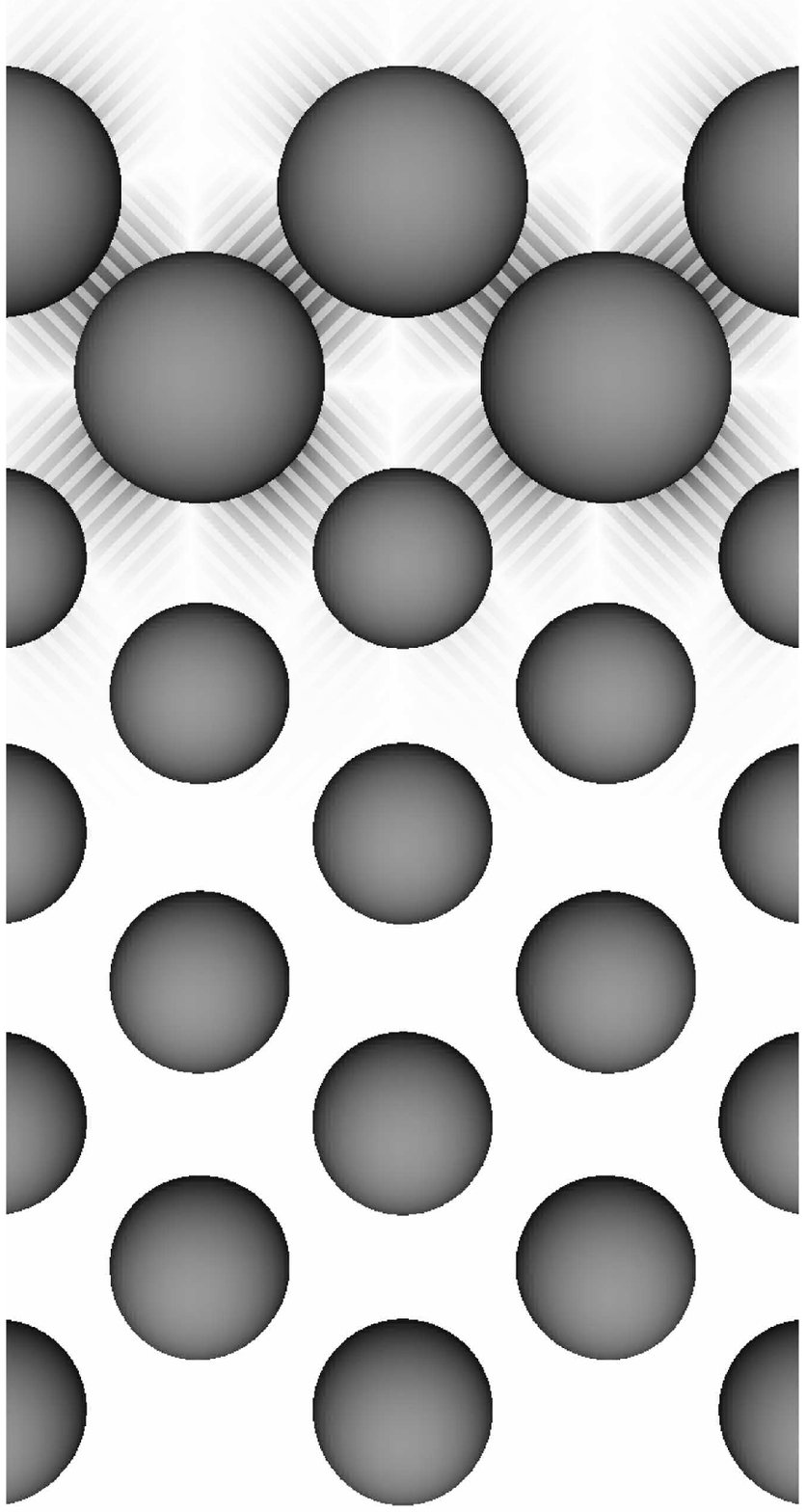}
\hspace{2em}
\includegraphics[clip=true]{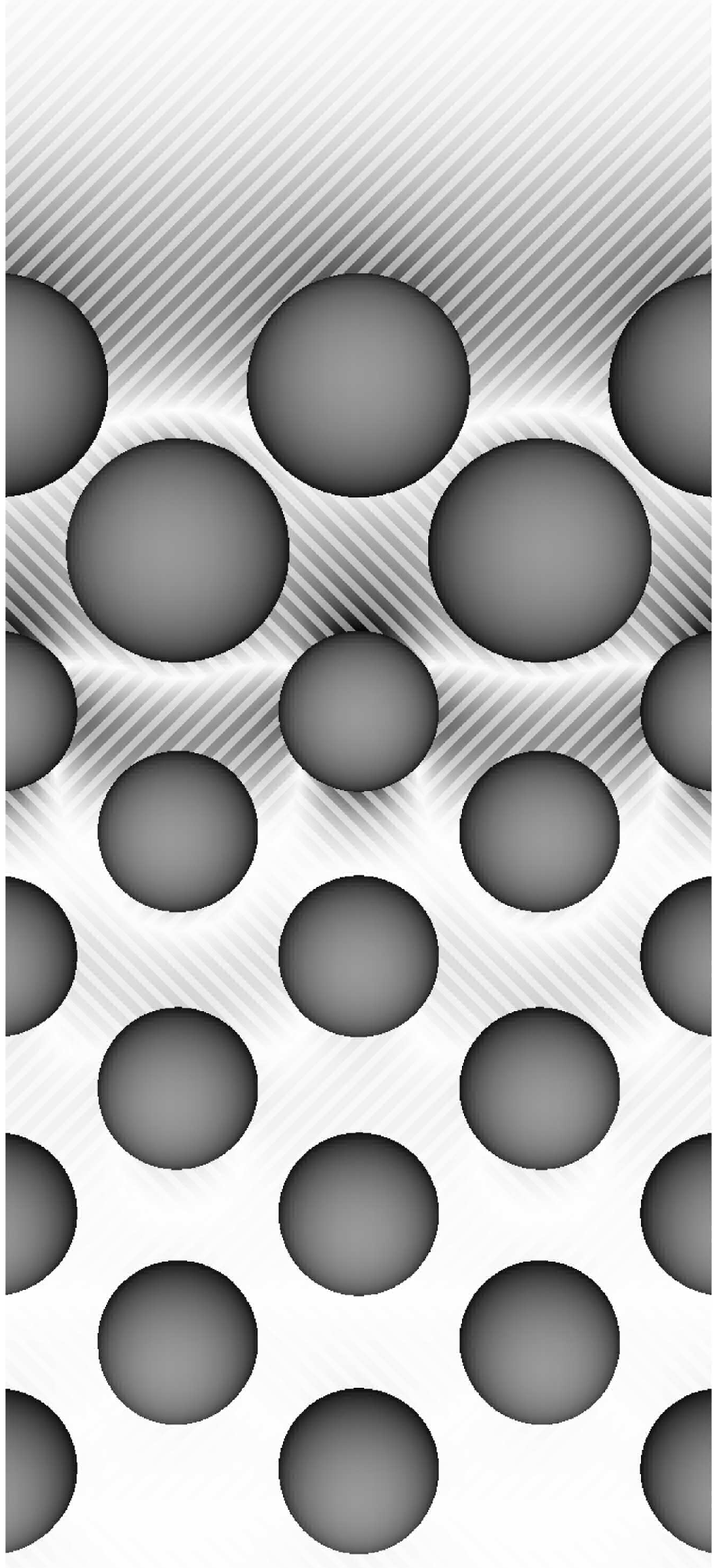}
}}
\caption{\label{eigenfunctions}
The wavefunctions of some electronic states in films of 2 ML Ag
on 7 layers of vanadium at $\bar{\Gamma}$ point, 
in the plane along $x=y$ and perpendicular to the surface.
The solid spheres hide the atomic cores, and the atomic pseudopotentials 
are defined at distances which lie inside the spheres.
The different shadings denote regions in which the wavefunctions have 
positive and negative values, respectively.
(a) Ag sp--V s QW state at 6.91~eV binding energy; (b) 
Ag sp with some Ag $\textrm{d}_{zz}$, hybridized with V s, at 5.45~eV; (c)
Ag $\textrm{d}_{xz}$ QW state at 4.59~eV; (d) Ag sp--V $\textrm{d}_{zz}$
QW state at 0.66~eV.
}
\end{figure*}

\begin{figure}[h]
\rotatebox{0}{
\resizebox{1.00\columnwidth}{!}{
\includegraphics[clip=true]{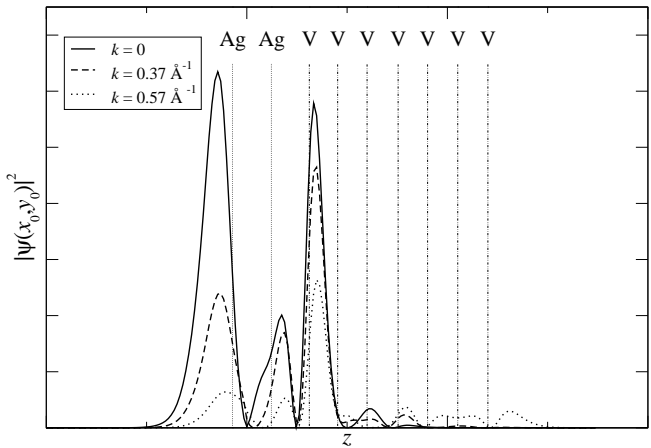}
}}
\caption{\label{ag2density}
The square of the wavefunction of the 2 ML Ag sp quantum well state as
a function of the coordinate $z$ perpendicular to the surface,
for three values of the wavevector $k_{||}$. The coordinates
$x_0=0$, $y_0=0.5a_{\mathrm V}$ are chosen so that the line does not
intersect any pseudopotential sphere used in the calculation.
}
\end{figure}

\subsection{Quantum well states and resonances}

In order to investigate into detail the character of quantum well states
and resonances, we have made intensity plots of the amplitudes of the
wavefunctions of some Kohn-Sham eigenstates. 
The wavefunctions shown in Fig.~\ref{eigenfunctions} are for 2 ML Ag 
films on V, but the results for 1 ML Ag films
are similar, although with fewer nodes along the $z$ direction.
We plotted the wavefunctions at the center of the surface 
Brillouin zone $\bar{\Gamma}$ where they can be expressed as a real
function by factoring out a constant complex phase.
The wavefunction of the sp QW state at $k_{||}=0$, $E=0.66$~eV, 
is particularly interesting. It is almost translationally invariant 
in the directions of the surface plane, with one node, as predicted by the 
classification of states in the simple potential well
models, e.~g. as shown in Fig.~10~(d) in Ref.~\onlinecite{Mil-02}. 
However, it has an unexpectedly large
amplitude on the first one or two vanadium layers, which is 
contributed by orbitals of $\textrm{d}_{zz}$ symmetry.
In fact, the (approximate) projection of the density of states onto atomic orbitals
of various symmetry shows that the wavefunction consists of
approximately half Ag sp and half V $\textrm{d}_{zz}$ orbitals.
Although large on the first V layer, the 
contribution of V orbitals decreases quickly and vanishes
within a couple of V layers, since these orbitals are pulled out 
from their `natural' energies (i.e.\ the respective vanadium 
d bands) and do not propagate into the vanadium bulk.

In Fig.~\ref{ag2density} we show the square of the wave function of
the 2~ML Ag QW state, for three values of the 
wavevector $k_{||}$ along the $\bar\Gamma$--$\bar{\mathrm M}$ direction. 
The functions are shown along the $z$ coordinate at a particular 
$x_0$ and $y_0$, chosen in such a way that the line does not
intersect any pseudopotential sphere used in our calculation.
This was done to avoid the strong variation of the wavefunctions close 
to the atomic cores, but apart from that the choice of $x_0$ and $y_0$ is not
significant, since the amplitudes of the wavefunctions are almost
translationally invariant, as can be seen in the density plot
of the QW state at $k_{||}=0$ in Fig.~\ref{eigenfunctions}~(d). 
Fig.~\ref{ag2density} confirms the quantum well state character of 
the wave function at $\bar\Gamma$, i.e.~$k_{||}=0$, where the 
{\it symmetry gap} is fully effective.
At $k_{||}=0.37$~\AA$^{-1}$, which is just inside the grey box in 
Fig.~\ref{2layers}, the wavefunction has a 
much decreased density outside the first Ag atom,
and extends up to the third and fourth V layer. Finally, at 
$k_{||}=0.57$~\AA$^{-1}$ the wavefunction is located mostly on the first two V layers, 
but extends through the rest of the slab and has a tail into the vacuum on 
the other side of our Ag/V structure. Thus the state has lost its QW character,
and can be more appropriately described as an interface resonance.

These results show that although simple well potential or phase accumulation 
models may successfully reproduce the energies of the QW states at 
the $\bar{\Gamma}$ point with an 
appropriate choice of parameters, only \textit{ab initio}
calculations can give accurate information about the degree of hybridization
with V orbitals, in particular away from the center of the surface 
Brillouin zone, which can be essential for photoemission
cross sections and other properties.

\subsection{Comparison with experiments}

\begin{figure}[h]
\rotatebox{0}{
\resizebox{1.00\columnwidth}{!}{
\includegraphics[clip=true]{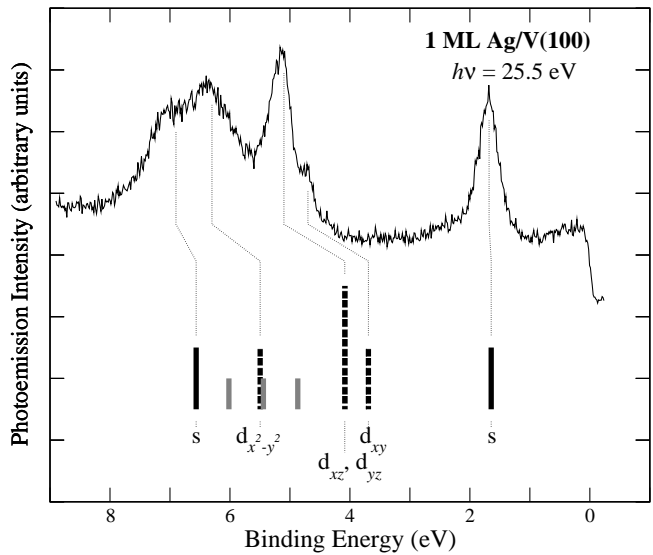}
}}
\caption{\label{PE-exp}
Experimental normal emission photoelectron spectra of 1~ML
Ag films deposited on V(100) surface, and the energy eigenvalues of 
Ag quantum well states and resonances obtained in DFT calculations
of 1~ML Ag on a seven layer V(100) substrate. Black bars: the QW 
states originating from Ag s orbitals, strongly hybridized with 
$\textrm d_{zz}$ orbitals on the underlying V atoms (the QW state at $E=1.65$~eV), 
and with the corresponding s orbitals (the QW state around
$E=6.56$~eV). Dashed bars: the Ag d QW states of various symmetries. 
Gray bars: the Ag s QW resonances, which lie within the vanadium s 
bands. See the discussion in the text.
}
\end{figure}

\begin{table}%[H] add [H] placement to break table across pages
\caption{\label{table}
Comparison of calculated and measured energies (in~eV, with respect 
to the Fermi level) of sp quantum well states
of 1 to 5 ML films of silver on vanadium.
}
\begin{ruledtabular}
\resizebox{0.95\columnwidth}{!}{
\begin{tabular}{ccc}
Number of Ag Layers  & Experiment\cite{Kra-01} & Calculation \\ \hline
1  & 1.65   & 1.65    \\
2  & 0.58   & 0.66     \\
3  & -- & 2.22   \\ 
4  & 1.43   & 1.61     \\
5  & 0.82   & 0.99   \\  
\end{tabular}
}
\end{ruledtabular}
\end{table}

In Fig. \ref{PE-exp} we have plotted the experimental spectra of normal
photoemission from a monolayer of Ag on V(100)~\cite{Mil-04} and calculated 
energies at the $\bar{\Gamma}$ point of electronic states localized on the silver 
adlayer. The experimental spectrum was measured at the ELETTRA synchrotron
in Trieste. The sample was at room temperature, and the 
energy resolution of the measurement was around 30--50~meV.~\cite{Kra-03b} The prominent
peak at 1.65~eV binding energy has been assigned to the Ag sp quantum well state,
and the one around 5.2~eV to the Ag d QW state. The FWHM of both peaks is
close to 400~meV. 

In Ref.~\onlinecite{Kra-01} a careful analysis of spectra taken in a similar experiment,
in which the energy resolution was around 10 meV and the temperature could
be varied, showed that at 60 K the 2~ML sp QW peak width is around 360 meV, which was 
decomposed into various contributions. Based on those results, we can attribute 
around 200~meV of the line width in Fig.~\ref{PE-exp} to the finite lifetime of the hole
(this contribution depends roughly quadratically upon the binding energy
of the QW state), 150~meV to impurity and phonon scattering, and the
remaining 30--50~meV to the experimental resolution.
In Ref.~\onlinecite{Kra-03b} it was found that a large part of the apparent width
of the d QW peak around 5.2 eV comes from the fact that it is a doublet, 
with the two components split by around 120 meV. The splitting is due to
the spin-orbit coupling, which is not included in our calculations. 
Another peak which can be attributed to a d QW state occurs at 4.7 eV binding 
energy. It is quite weak in the spectrum shown here but has a larger
intensity at other photon energies. It is a singlet, with a width around 
95~meV.\cite{Kra-03b}

The results of our calculations are shown as lines of various style
and height. The height indicates the degree of localization of the electronic
state on the silver adlayer, which gives a rough estimate of the expected
photoemission strength. Thus the state of s symmetry at 1.65~eV is shown 
with a full height line, since it lies in the symmetry gap of vanadium and 
cannot propagate into the bulk, although it has a strong component on
the first V layer, as already discussed for the 2 monolayer QW state.
Similarly, the Ag d states of $\textrm{d}_{xy}$ and $\textrm{d}_{x^2-y^2}$ 
symmetry at larger binding energies also form QW states due to symmetry 
mismatch with substrate bands, as do the $\textrm{d}_{xz}$, 
$\textrm{d}_{yz}$ states which are shown with a line of double height since 
they are degenerate at $k=0$. The states derived from Ag $\textrm{d}_{zz}$ 
orbitals, however, are shown as half-height grey lines, since they
readily hybridize with the s band of the vanadium substrate. These states 
have also a small admixture of Ag s orbitals. Finally, we have denoted the state
at 6.56~eV also by a black bar, i.e.\ as a QW state of s symmetry, because in
our calculation the wavefunction decreases approximately exponentially
on the deeper vanadium layers. However, our calculation with only seven
layers of vanadium underestimates the width of the s band of V, and a 
more appropriate description of this state may be that of a s resonance
at the bottom of the vanadium s band.

The calculated energy of the s QW state at $E=1.65$~eV agrees well with
the peak in the experimental spectrum. The energies of d quantum well 
states, however, are around 1~eV too high compared to the experimental peaks.
There is no reason for any shift of the experimental photoemission peaks
of the d states compared to sp, and this discrepancy appears to be 
a deficiency of the density functional calculation. It is indeed known
that local density functional calculations tend to place the d bands
of silver (and other noble metals as well) too close to the Fermi energy,
as discussed in Ref. \onlinecite{Fus-90} and references therein.

We obtain good agreement for the energies of the sp QW states at 
$\bar{\Gamma}$ point for thicker 
Ag films as well. In Table~\ref{table} we show the calculated energies 
of the highest state of sp symmetry, for films from 1 to 5 monolayers of Ag.
In all cases, except possibly for 3 ML, the energies fall in the
symmetry gap discussed earlier. The results are compared with
the values measured in normal photoemission.\cite{Kra-01} 
The agreement is very good, considering 
the finite width and the temperature dependence of the experimental peaks.
There is no direct measurement of the QW state in 3 ML Ag films, which are
probably structurally unstable.\cite{Kra-03a} However, 
in Ref.~\onlinecite{Mil-99} photoemission spectra have been reported
from Ag films on V(100) of nominal coverage of 2.5~ML, which presumably
consist of regions of 2~ML and 3~ML coverage. In Fig.~4 of that reference the broad
peak around $2.3$~eV, which is the upper edge of the vanadium s band, is
assigned to photoemission from a QW state or resonance associated with
the 3~ML regions, in good agreement with our calculated value.

The dispersion of Ag QW states can be compared with off-normal ARPES experiments.
In Fig.~\ref{2layers} we have marked by a grey box the $E-k_{||}$ region in the 
$\bar{\Gamma}$--$\bar{M}$ direction in which the angle resolved PES spectra
of 2 ML Ag films on vanadium have been reported.\cite{Kra-03c,Kra-05} Comparing the thick lines 
in our calculation to the density plot of the off-normal photoemission intensity
in Fig.~{6.14} of Ref.~\onlinecite{Kra-03c} and Fig.~2 of Ref.~\onlinecite{Kra-05} 
we see that the agreement is excellent. Both the 
quadratic dispersion of the QW state around the $\bar{\Gamma}$ point and the 
subsequent bending down, as well as the transfer of photoemission intensity to the 
vanadium-induced states which cross the Fermi level at finite $k_{||}$ are seen 
in the experimental spectra. It is interesting that our calculations find a quite
different behavior in the $\bar{\Gamma}$--$\bar{X}$ direction, where a state with
binding energy slightly larger than the QW state acquires a significant weight on
the surface atoms soon after leaving the 
$\bar{\Gamma}$ point of the Brillouin zone. This direction, however, has not 
been investigated by ARPES experiments. 

In Refs.~\onlinecite{Kra-03c,Kra-05,Kra-04}
standing wave patterns around surface defects have been observed in 
STM images taken on 1--5~ML thick Ag films on V(100) at room temperature. 
The defects were mostly point scatterers, and in some 
cases also surface steps. 
Such patterns are known to originate from the interference of 
electrons with predominantly two-dimensional character, usually surface 
states,\cite{Has-93} on localized potentials.
This is a rare example of standing waves in a room temperature
experiment, as most other observations have been done at low temperatures.
It indicates a strongly enhanced two-dimensionality of electronic states
in silver overlayers with energies around the Fermi level at large values 
of the wavevector $k_{||}$. In the unsupported silver monolayer, Fig.~\ref{monolayer}, 
the Fermi level is crossed by the s band at around $k_{||}=0.8$~\AA$^{-1}$.
For 1 ML and 2 ML Ag films on V(100), Figs.~\ref{1layer} and \ref{2layers},
states largely localized at the surface (thick lines) also appear
in this range of $k_{||}$. Strong hybridization with vanadium states 
makes the whole picture rather complicated, and a quantitative 
interpretation of the STM experiments is not possible at present.

\section{\label{conclusions} Conclusions}

Our calculations of the electronic states in few
monolayers thick silver films on V(100) substrate 
have shown that there is a strong localization of silver-derived states
within the Ag layer, due to the mismatch of the electronic structure of 
the two metals. 
The localization is particularly noticeable at the $\bar{\Gamma}$ point
of the surface Brillouin zone, where quantum well states of both 
s character (close to the Fermi level) and d character (at more than 3~eV
binding energy) appear on Ag films. States of $\textrm{d}_{zz}$ symmetry,
on the other hand, readily hybridize with vanadium s electrons, which are symmetry 
compatible. Hybridization effects also become more important away from
the center of the Brillouin zone.
We have obtained good agreement
with photoemission experimental spectra, both in the direction normal 
to the surface and with the available measurements in off-normal
directions. 

\begin{acknowledgments}

This work was supported by the Ministry of Science and Technology
of the Republic of Croatia under contract No. 0098001.
We acknowledge useful discussions with M.~Milun, P.~Pervan and M.~Kralj,
and their kind permission
to reproduce unpublished experimental spectra.
\end{acknowledgments}


\end{document}